# Secure Multi-Party Delegated Authorisation For Access and Sharing of Electronic Health Records

*Abstract*— Timely sharing of electronic health records (EHR) across providers is essential and significance in facilitating medical researches and prompt patients' care. With sharing, it is crucial that patients can control who can access their data and when, and guarantee the security and privacy of their data. In current literature, various system models, cryptographic techniques and access control mechanisms are proposed which requires patient's consent before sharing. However, they mostly focus on patient is available to authorize the access of the EHR upon requested. This is impractical given that the patient may not always be in a good state to provide this authorization, eg, being unconscious and requires immediate medical attention. To address this gap, this paper proposes an efficient and secure protocol for the pre-delegation of authorization to multi-party for the access of the EHR when patient is unavailable to do so. The solution adopts a novel approach to combine self-sovereign identity concepts and framework with secure multi-party computation to enable secure identity and authorization verification. Theoretical analysis showed that it increased the efficiency of the protocol and verification processes to ensure the security and privacy of patient's data.

## I. INTRODUCTION

Healthcare service providers and professionals operate various healthcare services at different locations. Usually, a user visits more than one healthcare professional, e.g., general practitioner, specialists, clinics, pharmacies, etc. for different needs. In a usual scenario, where users' health records that are issued by a healthcare provider are stored locally at the provider's data system as electronic health records (EHRs); all the management and maintenance of the data are on the provider side; only this provider is eligible to edit these records. On the aspect of availability and access to data, patients currently do not have a full, complete, let alone, comprehensive view of his or her medical history. Accessibility of health records by other healthcare service providers cannot be provided on a sufficiently prompt basis for patients and doctors to make correct and informed decisions on a timely basis. Insurance agents are unable to fully verify a client's (or claimant patient) full medical records before approving insurance claims or to facilitate checking and reviewing of complete medical information declarations made by clients during the purchase of insurance policies. Incorrect, incomplete and delayed medical information access could lead to patients paying incorrect premiums or making insurance claims that have exceeded or fall short of the actual reimbursable amounts.

For privacy and transparency issues, patients currently do not know how and to what extent have their medical records been utilised and the identities of parties that have access to their records. The level of appropriate privacy and transparency to patients' medical records may not have been adequately guaranteed or established. And on the fundamental matter of determining ownership of data, medical records (i.e., data) are created by the healthcare service provider upon a person registering as a patient with a clinical doctor or health service provider ( "HSP", e.g. Hospital). As such, the patient and medical practitioner or HSP should have joint ownership of the data. But authorisation to access should always be conferred on the patient since s/he is the rightful due data owner to decide how the data would be utilized.

In the aspect of entrusting medical records between the patient and the HSP, HSP is the data custodian. HSP is to ensure that the data is safe and secure, and to share the data only upon the patient's approval. Between or among HSPs, only after authorisation is obtained from the patient, the data can be shared only on a need-to-know basis and having based on mutual endorsement of a data sharing agreement. Additionally, the access of data is to be allowed only over a specified time frame to ensure its usage is on a need-to- know basis during the period of diagnosis or consultation. For health authorities to access data during emergency situations, eg, when patient is unconscious, this is only to be done after the patient has delegated the authorization or prior endorsed consent/authorise to the data sharing agreement

The use of blockchain technology ("BCT") has been advocated by research communities[1] in an attempt to overcome the challenges mentioned above and to address the gaps inherently in the healthcare industry. The BCT is a decentralized database and its properties of immutability, transparency and auditability, data provenance and availability can address some of the security concerns of the EHR sharing. However, it is unable to address adequately the security requirements pertaining to data confidentiality, privacy as well as access authorization of EHR.

In current literature, various system models and, cryptographic schemes and techniques are proposed. Majority of the reviewed literature requires data owner's (e.g., the patient) approval before sharing the EHR, but they have not taken into consideration the case whereby the patient is unavailable to perform the approval. There are often scenarios, as when the patient's health suddenly deteriorates, that require records be made available to HSP (eg specialists who could be remote) or other care givers who might not have initial access to the patient's health records. Existing authorization models follow a patient-centric approach where the EHR data authorization must be approved by the patient when required. This is not practical in some scenario and moreover the patient may not be in a state to provide this authorization when required. Hence there is a need to develop an authorization delegation mechanism whereby the patient can pre-authorizes the providers' access to his/her EHR in the event that s/he is certified as medically unfit to do so.

In regards to the issue on the control of data ownership, the notion of self-sovereign identity (SSI) has emerged in the past few years. SSI is a new paradigm of online identity management [2], whereby individuals and entities can manage their digital identity and identity-related information (i.e., identifiers, attributes and credentials, or other personal data) by storing them locally on their own devices (or remotely on a distributed network) and selectively grant access to this information to authorized third parties, without the need to refer to any trusted authority or intermediary operator to provide or validate these claims. SSI is a promising concepts that could be a means of confronting the challenge of sharing and securing sensitive medical information among healthcare parties, as well as ensuring patients maintain sovereignty over their data.

Thus, the focus of this paper is to propose an efficient and secure protocol for the pre-delegation of authorization[3, 4] to multi-party for the access of the EHR. The protocol facilitates the execution of the patient's pre-defined authorization to authorized parties, eg a panel of doctors can access the EHR when the patient is unconscious.

This paper's contributions are summarized as follows:
1) Proposes and designs an authorization security protocol that enables patients, as data owners, to pre-grant selected data requesters, example healthcare providers, access to and share their EHR.
2) Adopts a novel approach to combine self-sovereign identity (SSI) concepts and framework with secure multi-party computation (SMPC) to enable secure identity and authorisation verification in a decentralized setup. To the best of knowledge, this is the first research work that utilizes SSI, particularly Verifiable Credentials and Decentralized Identifier, for the purpose of granting authorisation using SMPC for verification and access to EHR.
3) Conducts a detailed security and privacy analysis of the security protocol using STRIDE [36] and LINDDUN [37].

The structure of this paper is organized as follows. Section II looks at related work and Section III provides the background for the components of the proposed solution. Section IV elaborates and discusses the system overview and design. Finally, Section V sets out the directions of future works and concludes the paper.

## II. RELATED WORK

Currently there are a number of researches conducted on the sharing of EHR using blockchain and different cryptographic schemes and access control mechanisms for secure sharing and access of EHR on a blockchain and cloud platforms. [5-10] proposed utilizing the blockchain platform as storage system for access-control model, protocols for authentication and sharing of healthcare data and access control for shared medical records in cloud repositories. [11-16] proposed a secure medical record sharing system using an attribute-based encryption and (multi-)signature scheme. [17, 18] proposed a blockchain based secure and privacy-preserving EHR sharing protocol using searchable encryption and conditional proxy re-encryption cryptographic schemes. [19] also uses searchable encryption but partitions patient's record into a hierarchical structure, each portion of which is encrypted with a corresponding key, thus enables patient to selectively distribute subkeys for decryption of various portions of the record. And [20] proposed MedChain, which combines blockchain, digest chain, and structured P2P network techniques to provide a session-based healthcare data-sharing scheme.

Majority of the above literature solution requires data owner's (e.g., the patient) to be available to approve before sharing the EHR, but only a few have taken into consideration the case whereby the patient is unavailable to perform the approval, e.g., in cases when he/she is unconscious in an emergency situation or mentally unfit to perform any tasks. [21] mentioned the use of an 'allowed list' for clinicians to access patient's data under emergency situation via prior one-time authentication from the patient. But as it is under an umbrella account of the HSP that links all clinicians (i.e. shared account), data security and privacy of the patient can be a major concerns, especially those not involved in the patient's medical consultations. [22] proposes the concept of using organisational structure roles to define entity-to-entity relationships and access rights based on functional roles and duties. This structure is used for authorisation management as well as access control. However, this way of access control is specific to a pre-defined organization structure and may not be aligned with the intention of the patient, the rightful data owner. [23] proposes a distributed system for delegation management using their eTRON enterprise security architecture that enables a patient to securely delegate access rights to her health records to someone s/he trusts. eTRON functions much like the SSI framework which has issuer to issue authorization token and this token is used for access to EHR. The solution requires hardware specific eTRON card with chip that stores the holder's identity. Unlike SSI, whose building blocks components and standards are defined by W3C, eTRON is more propriety which may have interoperability issues for wide deployment. [24] uses Attribute Based Encryption (ABE) and allows for delegated secure access of patient records. It similarly requires an organization structure of the entities or stakeholders of a medical organization and its patients to map out the access control rights base on the entities' attributes.

Self-Sovereign Identity (SSI), a decentralised technology for digital identity management, is a promising concept for handling health data. It could represent a step forward in empowering users, granting them control over their data [25]. [26] conducts a systematic literature review to investigate state-of-the-art measures based on SSI and Blockchain technologies for dealing with electronic health records (EHRs). It concludes SSI is still a novel subject and, even though adopting the principles of SSI could make patient-centric solutions more accurate, current healthcare research has neither adequately defined nor employed it in the health context.

The solution proposed by this paper combines SMPC scheme to delegate the authorization and adopts the SSI principles and framework to ensure the validity and verification of the identities, credentials and claims. To the best of knowledge, there is currently no related work on this approach.

## III. BACKGROUND

### A. Self-sovereign Identity

Self-sovereign identity (SSI) is a digital identity framework where an entity (an individual or an organization) owns its identity and controls the way it is shared in a decentralized setup, thus empowering the entity, granting it control over its data. Decentralized Identifier (DID) and Verifiable Claims /Credential (VC) are the essential building blocks of the SSI framework [2]. DID is a new type of identifier for verifiable, self-sovereign digital identity that is universally discoverable and interoperable across a range of systems and a standard defined by The World Wide Web Consortium (W3C)[27], analogous to a digital certificate issued by a certificate authority [33]. It is an URL (i.e., unique web addresses) associated with at least one pair of cryptographic keys: a public key & a private key. Together, the DID and public keys are published in the blockchain, and this "package" is called a DID document. A DID Document provides information on how to use t specific DID. For example, a DID Document can specify that a particular verification method (such as a cryptographic public key or pseudonymous biometric protocol) can be used for the purpose of authentication. Fig. 1 shows the DID data model. A DID by itself is only useful for the purpose of authentication. It becomes particularly useful when used in combination with verifiable claims or credentials (VC), another W3C standard, that can be used to make any number of attestations about a DID subject [28]. These attestations include credentials and certifications that grant the DID subject access rights or privileges. A verifiable claim contains the DID of its subject (e.g., a HSP), the attestation (access approval), and must be signed by the person or entity making the claim using the private keys associated with the claim issuer's DID (e.g., the patient). Verifiable claims are thus methods for trusted authorities (parties) to provably issue a certified credential associated to a particular DID to grant consent. It also guarantees privacy by enabling methods such as minimum/selective disclosure. Fig. 2 shows a typical structure of a VC.

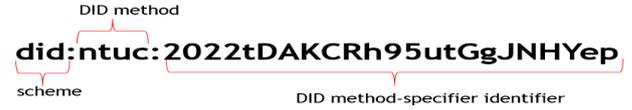

**Fig. 1.** An example DID data model. Method provides detail of where to fetch the DID and method-specifier identifier provides DID's unique identifier within the method.

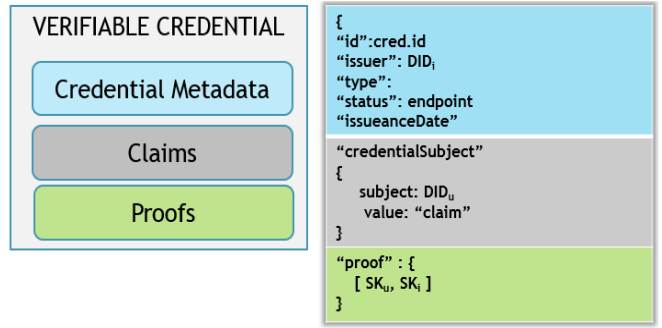

**Fig. 2.** A typical VC structure: Credential Metadata provides properties or attributes of the credential, Claims provides a statements about a subject and Proofs provides cryptographic signatures tied to private keys that prove the user sharing the VC is the subject of the information.

### B. Secure Multi-Party Computation (SMPC)

Secure multi-party computation (SMPC) protocol, such as oblivious transfer [29], homomorphic encryption (HE) [30] and the secret sharing scheme (SSS) [31], provides enhanced privacy, correctness and independence of inputs, and guarantees output delivery. It suits a distributed network like blockchain as it deals with security and trust issues in distributed environments. It is helpful in the scenarios whereby confidential data are to be shared across several organizations, across several sources and to run some kind of joint aggregation analysis or processing. Only specifically crafted shards of the data are exchange and every shard reveals nothing about the original data and it alone cannot be used to restore to the original. However the joint processing of shards is still possible to analyze the data. SSS is a form of multi-party computation, whereby a secret is divided into parts, giving each participant its own unique part. To reconstruct the original secret, a minimum number of parts, known as the threshold, is required. In the threshold scheme, this number ($t$) is less than the total number of parts ($n$), otherwise all participants are needed to reconstruct the original secret. The secret sharing scheme is defined as follows:

Let $\mathcal{P}$ be a set $\{P_1,…,P_n\}$ of $n$ entities, called participants, who take part in sending and receiving communications. An access structure on $\mathcal{P}$ is a collection $\mathcal{A}$ of subsets of $\mathcal{P}$. A subset $A \in \mathcal{A}$ is called an authorised set (of participants). Thus any set of participants that contains an authorised subset is authorised. In a monotone access structure, a minimal authorised subset is a subset $A \in \mathcal{A}$ such that $A \setminus \{a\} \notin \mathcal{A}$ for all $a \in A$, and a maximal unauthorised subset is a subset $A \notin \mathcal{A}$ such that $A \cup \{a\} \in \mathcal{A}$ for all $a \in \mathcal{P} \setminus A$.

For $i = 1,\ldots,n$, let $S_i$ denotes a set of elements, called shares corresponding to participant $P_i \in \mathcal{P}$. A secret sharing scheme for the key set $K$ on the set of participants $\mathcal{P}$ is a subset $\mathcal{D}$ of $K \times S_i \times \ldots \times S_n$ together with a probability distribution defined on $\mathcal{D}$. If $(k, s_1, \ldots, s_n) \in \mathcal{D}$ then we say that key $k$ is shared among the participants $P_1, \ldots, P_n$ who hold shares $s_1, \ldots, s_n$ respectively. The probability distribution on $\mathcal{D}$ induces a probability distribution on $K$ and each of $S_i$, $i = 1,\ldots,n$. The set $\mathcal{D}$ is a secret sharing scheme for $K$ with respect to the access structure $\mathcal{A}$ on $\mathcal{P}$ if

$$H(K \mid S_{i1},\ldots,S_{it}) = 0 \text{ iff } P_{i1},\ldots, P_{it} \in \mathcal{A}$$
$$> 0 \text{ iff } P_{i1},\ldots, P_{it} \notin \mathcal{A}$$

for all subsets $\{P_{i1},\ldots, P_{it}\} \subseteq \{1,\ldots, n\}$ and shares $s'_{i1},\ldots, s'_{it}$ where $s'_{ij} \in S_{ij}$ for $j = 1,\ldots, t$ there is a $k' \in K$ such that $k = k'$ for every $(k, s_1,\ldots, s_n) \in \mathcal{D}$ with $s_{ij} = s'_{ij}$ for $j = 1,\ldots,t$ if and only iff $P_{i1},\ldots, P_{it} \in \mathcal{A}$. We say that an authorised subset of participants $P_{i1},\ldots, P_{it}$ pool their shares $s'_{i1},\ldots, s'_{it}$ to get the key $k'$.

Secret sharing schemes defined on **n** participants, whose access structure consists of all sets of size of at least **t** are referred to as (**t, n**)-threshold schemes.

[32] proposes a solution to provide shared encryption (decryption) by applying the secret sharing techniques to the sharing of block cipher. That is either the encryption or the decryption of a message sent using that block cipher is a process to be distributed amongst a group of entities. They proposed 2 techniques, using cascading and XOR, for the composition of block ciphers. When an authorised group wish to encrypt a message or decrypt some ciphertext they cooperate by taking part in a protocol. This protocol enables them to perform the distributed computation of the cipher. This is an approach this paper adapts.

## IV. SYSTEM OVERVIEW AND DESIGN

### A. Solution Overview

The focus of the proposed solution is on delegation of authorization to HSPs or data requesters to access the patient's own EHR in the event that the patient is unconscious or mentally unfit to grant the approval in order for immediate medical care to proceed. Each of the participants in this ecosystem is issued with a DID, whether it is an individual (eg patient) or an entity (eg organization like HSP). The DID is also recorded on the blockchain (BC). HSP typically has a copy of the patient's EHR in their local system. Patient, as the data owner (DO) of the EHR, can request for his/her EHR to be accessible externally, eg via a cloud storage provider (CSP). HSP encrypts the EHR and stores it in CSP, the data custodian (DC). HSP provides DO with the secret key, ehr-id and DC identity. The secret key is required to decrypt and access the EHR and to ensure multi-party validation before the secret is revealed, multi-party computation is required. DO will generate the set of keys according to the number of authorized parties (**n**) and the minimum parties (**t**) needed to reveal the secret key. DO will encrypt the secret key with the set of keys. This set of keys is then split partially to the *n* parties whereby *t* parties will have all the set of keys to decrypt and derive the secret key.

In order to ensure the validity of the authorization process, one or more Notaries are identified as a witness. The Notary can be a lawyer or a trusted independent party. This is analogous to the Power of Attorney (Lasting Power of Attorney) process [38]. The set of keys is split among the *t* parties (Notaries and DC) and encrypted with their respective public keys which are recorded in their DIDs. For transparency and accessibility, the *ehr-id* and the encrypted key sets are recorded on the blockchain using DO generated *pseudoID*. DO can now issue a signed verifiable credential (VC) proving who are authorised to access to his/her EHR, with details of Notaries (witness/lawyer), DC, *ehr-id*, DO's *pseudoID*, expiry date and the encrypted secret key. By using a new *pseudoID* every time, DO's privacy can be protected. The VCs are cryptographically signed by the DO and issued to authorised parties, the data requesters (DRs). The VC provides DRs with the claim that DRs is authorized by DO for the access to the EHR identified by *ehr-id* and only VC has the link between DO's DID and *pseudoID*. When DR, holder of the VC, needs access to DO's EHR, he/she uses the VC and disclose the essential details to one of the Notaries to verify the authorisation. The Notary will decide depending on validity and expiry of the VC, and a check of the revocation list. Once Notary validated, DR will need to work together with DC and Notary to decrypt and retrieve the encrypted secret key in the VC. Since only DC knows the storage location of DO's EHR linked to the *ehr-id*, it will provide a link for DR to access the encrypted EHR. DR can impose a time period for the access – eg availability of the link. DR can downloaded the encrypted EHR and decrypt it with the secret key to access the EHR content.

In a SSI model, DO is the issuer, Notary and DC are the verifier and DR is the holder of the VC. The VCs are presented via Verifiable Presentation (VP) and with VPs, the holders (for our case DR) can freely choose which information (from underlying VCs) they include in a Verifiable Presentation and thus, share with a relying party. This is one of selective disclosure feature of SSI solution. Additional access rights and attributes can be defined in the VC to provide more fine-grained access control of the EHR content.

### B. Functional Flow

The functional flows are broken down into 3 parts; namely, Secure storing of EHR, Delegation of authorizations to DRs and Secure Access to EHR. It is assumed that all parties' DIDs are recorded and verifiable on blockchain (BC).

#### 1) Secure storing of EHR

Patient has consulted a medical practitioner from a HSP and his/her EHR is recorded in HSP's private data store. Patient requests the EHR to be made available to him/her.

a) HSP retrieves patient's (DO) EHR from its private data store and encrypts it with **sk** before uploading to a public cloud storage provider [34] (DC). HSP is provided with **ehr-id** which is used to locate the encrypted EHR (**$EHR_{sk}$**) in DC's data store. The DC is assumed to be semi-honest.

b) HSP encrypts **sk** and **ehr id** with DO's public key (which is within DO's **DID**) and sends to DO through secure channel, eg Transport Layer Security (TLS).

c) DO decrypts with its private key (from DO's **DID** in its personal wallet) and store the **sk** and **ehr id.**

#### 2) Delegation of authorization to DRs

DO wishes to pre-assign parties with the authorization to access her/his EHR in the event s/he is unfit to do so.
a) DO identifies the parties (DRs) which it would like to grant the authorisation to access its EHR.
b) DO generates a set of keys (depending on $n$:number of Notaries + DC) and $t$: minimum parties needed – eg [n=3, t=2]) and a nonce, $r$, and encrypts $r$ with the keys and XOR with $sk$ to derive, *cipherKey*.
c) DO splits the keys each for DC and Notaries and encrypts them using their public keys to derive, *encryptedKeys$_i$*, $i$ is the party index.
d) DO generates a pseudoID for BC and records *encryptedKeys$_i$*, its *pseudoID* and *ehr-id* on BC.
e) DO generates for each DR a verifiable credential (*VC*) and input the *DIDs* of the DR, Notaries and DC, *cipherKey, pseudoID, ehr-id* and *VC* expiry date. DO digitally signs each *VC* using its private key, and encrypts the data using each DR's public key.
f) DO issues the *VC*s to the DRs through secure channel.
g) DRs decrypts the *VC*s with their private keys and store the *VC*s in their repository.

### 3) Secure Access to EHR

In the event that DO is unconscious and unable to authorise the access to his/her EHR:
a) DR needs access to DO's EHR.
b) DR retrieves *VC* from its repository and reads DO's *pseudoID*, *cipherKey*, Notaries and DC's *DIDs*.
c) DR extracts nonce, $r$, from *cipherKey*.
d) DR contacts a Notary and presents the *VC* as a verification presentation, disclosing only the relevant details –DO's signature, $r$, *pseudoID* and *ehr-id*.
e) Notary verifies DO's state, availabilty – an offline process.
f) If DO is available, Notary will seek DO's approval, else access is granted based on authenticity and expiry of *VC* as well as a check against the revocation list available on BC.
g) If granted, Notary reads from BC its keys (*encryptedKeys*) based on DO's *pseudoID* and *ehr-id*. Notary decrypts its keys (*encryptedKeys*) with its private key and encrypts $r$ with the keys to derive *partialCipher$_i$*. *partialCipher$_i$* is returned to DR.
h) DR similarly presents *VC* to DC with details of the DO's signature, Notary *DID, pseudoID, r,* and *ehr-id.*.
i) DC verifies the *VC* and optionally with Notary.
j) DC searches BC based on *pseudoID* and *ehr-id*, reads and decrypts its keys (*encryptedKeys*) with its private key and encrypts $r$ with the keys to derive *partialCipher$_i$*. *partialCipher$_i$* is returned to DR together with the link to download the encrypted EHR, *EHR$_{sk}$*.
k) DR uses XOR of all recieived *partialCipher$_i$* with *cipherKey* to derive *sk* which is used to decrypt *EHR$_{sk}$* and retrieve the EHR records.

The secret key that encrypts the EHR is derived via SMPC: DR extracts nonce in *cipherSK* in VC, gets DC and Notary to encrypt the nonce with keys they each have, and xor together with *cipherSK* to derive the secret key to decrypt the encrypted EHR.

### C. Threat Modelling

Blockchain only addresses a portion of the desired security requirements, in terms of transparency, integrity, availability and thus a certain level of trust the blockchain technology provides. However the other security requirements are also needed to be addressed, namely; Confidentiality, Privacy and Improved level of Trust.

To put it into a better perspective the security and privacy threats of the proposed solution, threat modeling is performed using a data flow diagram (Fig. 3) together with a security analysis (TABLE I) and a privacy analysis (TABLE II) to illustrate the threats exhibited by the system functions. The data flow diagram in Fig. 3 illustrates the data flow as was elaborated in the functional flow in earlier section. The tables show likely threats face by each of the data flow elements. A discussion of the threats and how the proposed solution

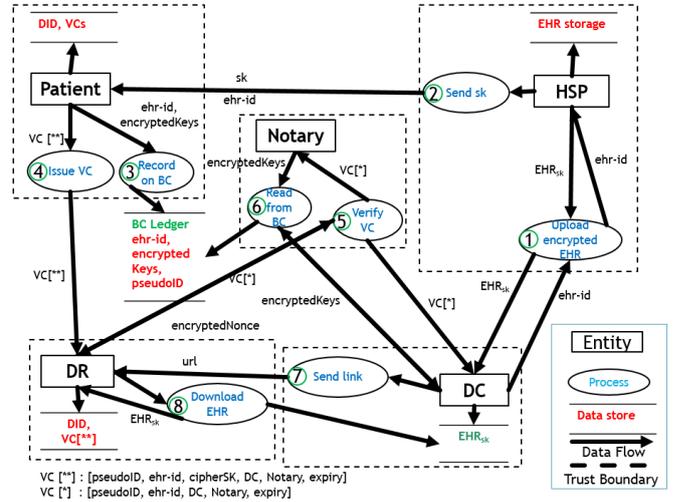

**Fig. 3.** Data Flow Diagram for a typical system for storing and sharing EHR using blockchain

| type | DFD elements | S | T | R | I | D | E |
|---|---|---|---|---|---|---|---|
| | Security Analysis | | | | | | |
| DS | Patient personal data store | | | | | | |
| DS | DR private data store | | | | ✓ | | |
| DS | DC data store (public cloud) | ✓ | ✓ | ✓ | ✓ | ✓ | ✓ |
| DS | HSP data store (private cloud) | | | | | | |
| Proc | Upload encrypted EHR | ✓ | ✓ | ✓ | ✓ | | |
| Proc | Send sk (secret key) | ✓ | ✓ | ✓ | | | |
| Proc | Record on BC | | ✓ | ✓ | ✓ | | |
| Proc | Issue VC | ✓ | | ✓ | ✓ | | |
| Proc | Verify VC | ✓ | | ✓ | ✓ | | ✓ |
| Proc | Read from BC | | | | ✓ | ✓ | |
| Proc | Send Link | ✓ | ✓ | ✓ | ✓ | | |
| Proc | Download EHR | | | ✓ | | ✓ | ✓ |
| Entity | DR (data requester) | | | | ✓ | | |
| Entity | DC (data custodian) | | ✓ | | ✓ | | ✓ |
| Entity | HSP (healthservice provider) | | | | ✓ | | |

addresses them follows:

*1) Security Analysis*

A security analysis is performed using STRIDE. STRIDE [36] is a threat model developed by Praerit Garg and Loren Kohnfelder at Microsoft for identifying security threats. It provides a mnemonic for security threats in six categories, namely; Spoofing, Tampering, Repudiation, Information disclosure, Denial of Service and Elevation of privileges. The security threats present in the system according to six categories are discussed. In addition, the collusion and key management threats are also discussed:

- **Spoofing**:
  (i) The identities of the participants are recorded on BC in the form of DIDs which stores the public keys which can verify the participant identity and signature.
  (ii) The corresponding private key is safely stored in the participant's wallet or repository which can be retrieved for cryptographic operations and generating digital signature.
  *Threat*: A likely spoofing of identity threat can be validation and verification of the DID prior to recording onto BC which relates to the implementation of consensus protocol.

- **Tampering:**
  (i) The VC is digitally signed by the issuer which ensures the integrity and authenticity of the signed content since it requires the issuer's private key to do so.
  (ii) DO's EHR is encrypted with secret key known only to him/herself and HSP. One additional measure is for DO to digitally sign the secret key before encrypting with a set of keys.
  (iii) DC will not be able to tamper with the EHR since it is encrypted and it is non-beneficial to itself to provide a tampered link to DR for downloading the encrypted EHR.

- **Repudiation:**
  (i) The interactions between the participants are direct and verification of identities are immediate, thus the participants cannot deny the interaction and actions taken.
  (ii) VC is digitally signed and issued by DO to DRs.
  (iii) Only DR knows how to derive the secret key to decrypt the data downloaded from the link provided by DC.

- **Information Disclosure:**
  (i) Only pseudoID and ehr-id are recorded on BC.
  (ii) DR can selectively disclose the need to know information when requesting for verification.
  (iii) DC cannot disclose any content of DO's encrypted EHR other than its storage location.
  *Threat*: Only threat is HSP and DR's revelation of EHR content after retrieving clear content. This can be mitigated through law abiding policies like NDA or code of ethics.

- **Denial Of Service:**
  (i) Except for DC which holds the location of EHR, all contents are on BC which is decentralized and available. However, this threat should be already mitigated by DC as with most cloud storage providers.

- **Elevation of Privileges:**
  (i) VC is only issued to the authorized DRs with details also recorded on the VC. Unauthorised DR cannot use the VC as its own or download the encrypted EHR.
  (ii) DC would have mitigated this risk as part of its security posture.
  (iii) Mutli-parties are involved with Notary to start off with the verification to proceed, and DC to finally grant access to EHR. Possible threat is collusion.

- **Collusion:**
  (i) Notary with DR – They do not know the other partial keys held by DC and the location of the EHR.
  (ii) DR with DC – They do not know the other partial keys held by Notary
  *Threat*: Notary with DR and DC – This is the only likely collusion threat that requires all 3 parties.

- **Key management:**
  (i) The set of keys used for SMPC are encrypted and stored on BC with link to DO pseudoID and ehr-id. This removed the need to store in a secure location for the participants (Notaries and DC) and can read from BC and extract the keys using their private keys.
  (ii) DO need only store the secret key to the encrypted EHR and optionally the encrypted nonce using the set of keys.

*2) Privacy analysis*

A privacy analysis is performed using LINDDUN. LINDDUN [37] is a privacy threat modeling methodology that supports analysts in systematically eliciting and mitigating privacy threats in software architectures. It provides a mnemonic for privacy threats in seven categories, namely; Linkability, Identifiability, Non-repudiation, Detectability, Disclosure of information, Unawareness, Non-

TABLE II      PRIVACY ANALYSIS USING LINDDUN

| type | DFD elements | L | I | N | D | D | U | N |
|---|---|---|---|---|---|---|---|---|
| DS | Patient personal data store | | | | | | | |
| DS | DR private data store | | | | | | | |
| DS | DC data store (public cloud) | ✓ | ✓ | ✓ | ✓ | ✓ | ✓ | ✓ |
| DS | HSP data store (private cloud) | | | | | | | ✓ |
| Proc | Upload encrypted EHR | ✓ | ✓ | ✓ | ✓ | | ✓ | |
| Proc | Send sk (secret key) | | | ✓ | | | ✓ | |
| Proc | Record on BC | ✓ | ✓ | ✓ | ✓ | | ✓ | ✓ |
| Proc | Issue VC | | | ✓ | ✓ | | ✓ | |
| Proc | Verify VC | ✓ | ✓ | ✓ | | | | |
| Proc | Read from BC | ✓ | ✓ | | | ✓ | | |
| Proc | Send Link | ✓ | ✓ | ✓ | ✓ | | ✓ | |
| Proc | Download EHR | | | ✓ | ✓ | ✓ | ✓ | |
| Entity | DR (data requester) | | | | | | | ✓ |
| Entity | DC (data custodian) | | | | | | | ✓ |
| Entity | HSP (healthcare provider) | | | | | | | ✓ |

compliance. The privacy threats present in the system according to seven categories are:

(i) **Linkability** (able to link items of interest to know the identity of the data subject(s) involved):
  (i) Only DO's pseudoID is used on BC and there is no link between it and DO's identity (DID). The only link is found on the VC which is stored internally– this link is required to identify DO's ehr-id block on BC.
  (ii) The EHR is not publicly available and only DC can grant its access. In addition, DC only knows the ehr-id and not the content of the encrypted EHR.

- **Identifiability** (to able to identify a data subject from a set of data subjects through an item of interest):
  (i) PseudoID on BC is not able to identify who is the DO. The pseudoID is generated new for every VC.
  (ii) There is no content on DC that can identify who the DO is.

- **Non-repudiation** (from data owner's perspective of able to deny a claim):
  (i) The VC stored all required details to identify DO and its signature for his/her authorisation for DRs.
  (ii) Only DR knows how to derive the secret key to decrypt the data downloaded from the link provided by DC.
  (iii) The interactions of DR with Notary and DC are also recorded on BC for reference.

- **Detectability** (able to distinguish whether an item of interest about a data subject exists, regardless of being able to read the contents itself):
  (i) With only pseudoID and ehr-id recorded on BC, there is no trace to detect it is DO's EHR but only an EHR is stored. The pseudoID will not be the same for the same DO.
  (ii) The EHR is not publicly available and only DC can grant its access. In addition, DC only knows the ehr-id and not the content of the encrypted EHR.

- **Disclosure of Information:**
  (i) BC only stored pseudoID, ehr-id and encrypted keys with no other details.
  _Threat_: Only risk is DR's disclosure of information after decrypting EHR – this is beyond any control since EHR is already in clear.

- **Unawareness** (unaware of the actions done on the one's (data subject) personal data):
  (i) DRs need to request Notary in order to access EHR. Notary will notify DO for any access to his/her data.
  _Threat_: Notary may choose not to notify DO. As the solution is on assumption that DO is unavailable to grant access, DO can detect the access only via the transaction recorded on BC.

- **Non-compliance** (action done on personal data that is not compliant with legislation, regulation, and/or policy.):
  (i) As a participants of this healthcare ecosystem, each participant will have accepted the term and conditions of use and to abide to the medical code of ethics, policies, laws and regulations specific to its country.

To further analyse the disclosure of information and the information available to each participants and non-participants, a 'who knows what' table is shown in TABLE III.

TABLE III. WHO KNOWS WHAT

| | DO | HSP | DR | Notary (Trusted) | DC (semi-honest) | Outsider | Remarks |
|---|---|---|---|---|---|---|---|
| Secret key to encrypt EHR | ✓ | ✓ | | | | | Only DO and HSP |
| **DO DID | ✓ | ✓ | ✓ | ✓ | ✓ | ✓ | On BC |
| DO pseudoID | ✓ | ✓ | ✓ | ✓ | ✓ | ✓ | On BC |
| DO's DID-pseudoID link | ✓ | | ✓ | ✓ | ✓ | | On its VC |
| cipherSK | ✓ | | ✓ | | | | On its VC |
| ehr-id, encrypted keys | ✓ | ✓ | ✓ | ✓ | ✓ | ✓ | On BC |
| Who is Notary | ✓ | | ✓ | ✓ | ✓ | | On VC |
| Who is DC | ✓ | ✓ | ✓ | | ✓ | | On VC |
| Who are on DO's authorisation list | ✓ | | ✓ | ✓ | | | On VC:Only its own, Notary will have full list |
| EHR storage location/link | | | | | ✓ | | |
| DO's keys to generate cipherSK | ✓ | | | ✓* | ✓* | | *Partial only |

**All DIDs are on blockchain to serve as identity proof with its public keys in a DID Document

This provides further analysis of the privacy protection of the proposed solution and any information available to any participants is on a need-to-know basis. Utilising the selective disclosure of VC, DR need not provide the encrypted secret key, *cipherSK*, to Notary or DC when requesting them to verify the VC. DO's pseudoID used on BC will not be linkable to DO's DID unless the parties are part of the process since the partial keys are recorded on BC and DO's DID need to be verified.

The interactions between parties are also on a need basis. DO is required to interact with HSP to locate its EHR in DC and DRs when issuing the VCs to them. DR will need to interact with DO to receive the VC and with Notary and DC for verification of the VC and receive the ***partialCipher$_i$***.

## V. CONCLUSION AND FUTURE WORK

Timely sharing of electronic health records (EHR) across providers is essential and has great positive significance in facilitating the medical research of diseases and doctors' diagnosis for prompt patients' care. It is also important for patient, as the rightly data owner, to have full control of his/her EHR and grant the access to the EHR accordingly. Current researches have looked into different cryptographhics techniques and access control to ensure the security and privacy of the shared EHR. However, it is also essential to address the authorization concerns when patient is unavailable, eg unconscious in an emergency, to grant the consent for the access of the EHR for immediate medical attention. This paper proposed and designed an authorization security protocol that enables patients, as data owners, to pre-grant selected data requesters access to and share their EHR. The design adopted the Self-Sovereign Identity (SSI) concepts and framework, particularly Decentralized Identifier (DID) and Verifiable Claim/Credential for authentication and authorization respectively and combined with secure multi-party computation (SMPC) to enable secure identity and authorisation verification in sharing of the EHR and protect patient's privacy through selective disclosure. A security and privacy analysis were conducted on the protocol and discussed.

An implementation of the protocol is on-going. A suitable SSI frameworks [2] will be adopted and a SMPC implementation assessing the XOR and cascade approach [32] will be conducted. The SMPC implementation will be integrated into the eventual SSI framework and blockchain platform with sample medical data [35] for testing. In addition, access rights and attributes can be further defined in the VC to provide more fine-grained access control of the EHR content. The proposed solution and implementation can also be explored and adapted for other domains, eg the processing and execution of Lasting Power of Attorney (LPA) and will.